\documentclass[a4paper,twocolumn]{article}
\usepackage{epsfig,cite}
\begin{document}
\twocolumn[
\title{\textbf{Conversion efficiency\\
in a resonant Josephson effect mixer}}
\author{
Valentina S. U. Fazio\\
\textit{\small{Department of Physics f9a, Chalmers University of Technology 
and G\"oteborg University}}\\
\textit{\small{S-41296 G\"oteborg, Sweden}}
}
\date{}
\maketitle
\begin{center}
\begin{minipage}{0.85\textwidth}
%
%\begin{abstract}
The RCSJ (Resistor-Capacitor Shunted Junction)
model provides us an analytical formula for the 
conversion efficiency for a resonant Josephson effect mixer.
An external rf signal is applied to a long Josephson junction
operating in the fluxon oscillating regime.
The sine-Gordon equation is used to investigate the dynamics
of the phase-locking between the fluxon oscillations and the 
external signal.
We obtain that it is possible to have conversion efficiency
greater than unity.  
%\end{abstract}
\bigskip
\end{minipage}
\end{center}
]

\section*{Introduction}
A Josephson junction having one physical dimension larger
than the Josephson penetration depth can support periodic motion of
magnetic flux quanta\cite{BarPat, VanDuz}.
The evidence for this is the existence of singularities in the current-voltage (I-V)
characteristic of the junction, known as zero-field steps (ZFS).
Thus, a long Josephson junction (LJJ) dc biased on a ZFS emits electromagnetic
radiation whose frequency is proportional to the
geometric length of the junction.
Due to the low dynamic resistance of a ZFS, this radiation has a very narrow
linewidth\cite{CirRotJul95} and it is very 
interesting for potential practical applications as
local oscillator for a superconducting mixer:
the narrow linewidth can reduce the mixer-noise very much.

When an external signal is applied to the junction a new singularity in the
I-V curve appears, known as phase-locking step: the junction is
said to be in resonance with the external signal. 
Phase-locking of a long Josephson junction (LJJ) to an external rf drive
has been investigate in several publications by numerical simulation, theory,
and experiments.
Already, theoretical results have shown that some of the observed
experimental and numerical features can be reproduced by means of
analytical approaches, based on sine-Gordon fluxon dynamics.

In this paper we will also make use of the
Resistor-Capacitor Shunted Junction (RCSJ) model.
With this model conversion efficiencies larger than unity,
and in very good agreement with the experiments, have been found for
mixers build with
small Josephson junctions\cite{TaurThesis, TauClaJan74}
but it was never applied to LJJ.

The purpose of this paper is to apply the RCSJ model to a resonant
Josephson effect mixer to obtain an analytical formula
for the conversion efficiency. 

\section*{Conversion gain in a Josephson effect mixer}
The conversion efficiency, $G$, 
for a Josephson effect mixer
is defined as\cite{TaurThesis, TauClaJan74}:
\begin{equation}
G=\frac{P_{\mathrm{if}}}{P_{\mathrm{s}}},
\label{uno}
\end{equation}
where $P_{\mathrm{if}}$ is the power delivered to the room-temperature
amplifier and $P_{\mathrm{s}}$ is the power from the small signal
source.
In the RCSJ model $G$ can be conveniently written in the form:
\begin{equation}
G=C_{\mathrm{if}} \frac{R_{\mathrm{dyn}}}{R} \chi^{2},
\label{due}
\end{equation}
where $C_{\mathrm{if}}$ is the output coupling efficiency
(we will only consider the ideal case, $C_{\mathrm{if}}=1$), 
$R_{\mathrm{dyn}}$ is the
inverse slope of the I-V curve at the bias point, $R$ is the shunt
resistance of the junction, and $\chi$ is a 
dimensionless parameter\cite{TaurThesis, TauClaJan74}:
\begin{equation}
\chi= \frac{\partial (\Delta I/I_{\mathrm{c}}) }
{\partial \left[ \left( P_{\mathrm{rf}}/R 
{I_{\mathrm{c}}}^2 \right)^{1/2} \right]}.
\label{3}
\end{equation}
In equation (\ref{3}) $\Delta I$ is the height of the 
rf induced step on the I-V curve
in presence of the $P_{\mathrm{rf}}$ power, and 
$I_{\mathrm{c}}$ is the critical current of the
junction. 
Since $P_{\mathrm{rf}}={I_{\mathrm{rf}}}^2 R$ and 
$I_{\mathrm{c}}$ is a constant, we find:
\begin{equation}
\chi=\frac{\partial (\Delta I)}{\partial I_{\mathrm{rf}}}.
\label{quattro}
\end{equation}
Inserting (\ref{quattro}) in (\ref{due}) 
we can easily calculate the values of $G$ once we have estimated 
$R_{\mathrm{dyn}}$ and $\Delta I$ from the I-V curve of the junction.

We want to apply this theory to a resonant Josephson effect mixer
and demonstrate how it is possible to have conversion efficiency
larger that unity.

\section*{Dynamics of a long Josephson junction}
To modelling a long Josephson junction we consider
the forced sine-Gordon
system\cite{BarPat, VanDuz}:
\begin{equation}
\phi_{tt}- \phi_{xx} + \sin \phi + \alpha \phi_{t} = 
\gamma + \eta \sin \omega t,
\label{cinque}
\end{equation}
with the boundary conditions
\begin{equation}
\phi_{x}(0,t) = \phi_{x}(l,t) =0.
\label{sei}
\end{equation}
In the (\ref{cinque}) and (\ref{sei}) the length is normalised
to the Josephson penetration depth, $\lambda_{\mathrm{J}}$,
the time is normalised to the inverse of the Josephson plasma 
frequency, $\Omega_{\mathrm{j}}$, and the currents are normalised to the 
critical current, $I_{\mathrm{c}}$.

In equation (\ref{cinque}) 
the quantity $1/\alpha$ represents the Ohmic resistance 
of the junction and
$\gamma$ represents the bias current.
$\eta$ is the rf current
and $\omega$ is the 
frequency of the rf signal. 
The junction is uniformly rf-driven by the term $\eta \sin \omega t$.
This can be experimentally achieved in the overlap-geometry.
In equation (\ref{sei}) $l$ is the
normalised length of the junction.

For $\eta=0$ the system (\ref{cinque})-(\ref{sei}) 
%and with appropriate
%choices of the parameters $\gamma$, $\alpha$, and of the initial conditions,
exhibits fluxon oscillations over the spatial interval of length $l$,
whose frequency is the inverse of the time that the fluxon spends to
cover a distance $2\,l$.  
The I-V curve of the junction
will show the characteristic Zero Field Steps.

For $\eta \neq 0$ the system (\ref{cinque})-(\ref{sei}) shows current
steps due to the interaction between the external radiation and
the internal oscillations.
The height of the step $\Delta I$
for a given value of $\eta$ is called \textit{phase-locking
range}.
A very important feature of the system (\ref{cinque})-(\ref{sei})
is that the dependence of the phase-locking range upon the 
externally applied rf-current amplitude, $\eta$, is linear. 
If $\Delta I$ is the phase-locking range, we have that\cite{CirRotJul95}
\begin{equation}
\Delta I = \frac{1}{\sqrt{2}}\, \alpha\, \eta\, R_{\mathrm{p}},
\label{sette}
\end{equation}
where 
\begin{displaymath}
\frac{1}{R_{\mathrm{p}}} = \frac{\partial I(\gamma, \eta=0)}{\partial V} 
\end{displaymath}
is the inverse of the slope of the I-V curve
without external radiation applied, i.e. the slope of the Zero
Field Step, at the point where the phase-locking step appears.

%---------------------------------------------------------
\begin{figure}[t]
\epsfig{file=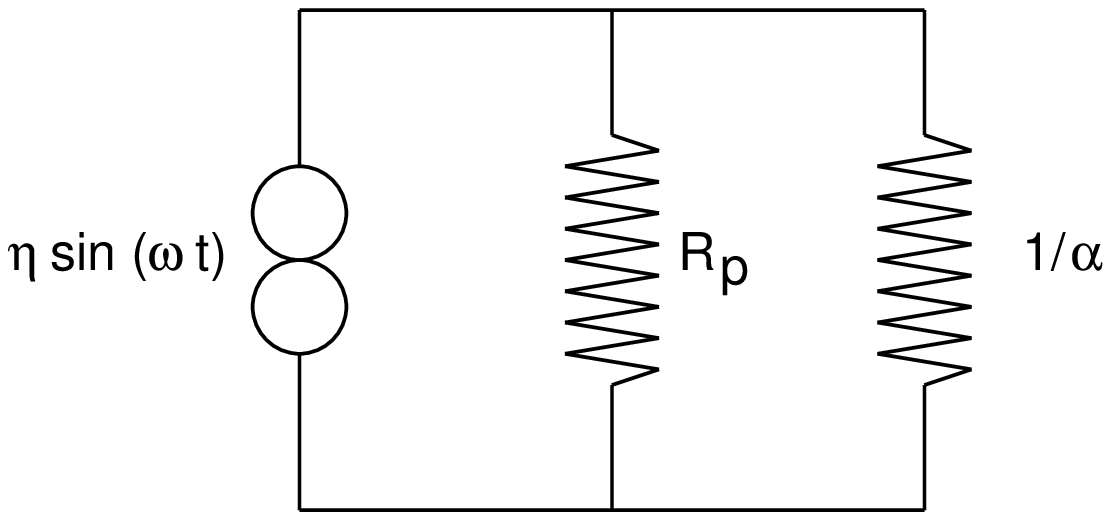,width=0.47\textwidth}
\vspace{7mm}
\begin{center}
\large{$R_{\mathrm{p}} \ll \frac{1}{\alpha}$} \\
$\Downarrow$
\end{center}
\epsfig{file=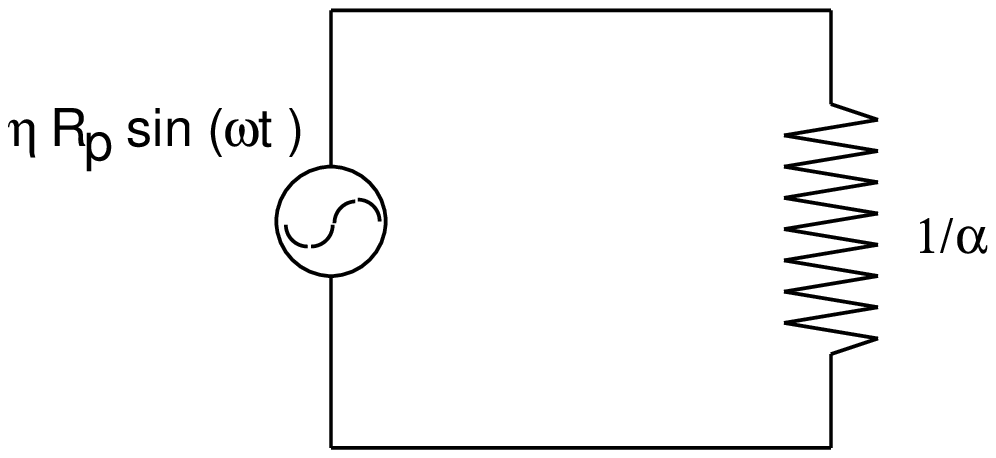,width=0.47\textwidth}
\caption{
\small{Circuit modelling of the driven Josephson junction
\protect{\cite{CirRotJul95}}.
Since $R_{\mathrm{p}}$, the dynamical resistance of the junction biased on 
a Zero Field Step, is usually much smaller than the Ohmic impedance
of the tunnel junction, all the current will flow
through $R_{\mathrm{p}}$. Then we can approximate the voltage source
$\eta \sin (\omega t)$ applied to the resistor $R_{\mathrm{p}}$
with a current source $R_{\mathrm{p}} \eta \sin(\omega t)$.}
} 
\label{modCirillo}
\end{figure}
%---------------------------------------------------------

The basic idea beyond the equation (\ref{sette}) is 
\cite{CirRotJul95}: A long
junction can be viewed (see Figure \ref{modCirillo}) as the parallel
combination of the dynamical resistance of the junction when biased
on a Zero Field Step, $R_{\mathrm{p}}$, and the Ohmic resistance $1/\alpha$.
Since usually we have that
\begin{displaymath}
R_{\mathrm{p}} \ll \frac{1}{\alpha},
\end{displaymath}
the parallel combination of these two resistances and the \textit{current
generator}, $\eta \,  \sin (\omega t)$, can be approximated
by a \textit{voltage generator} of amplitude $\eta \, 
R_{\mathrm{p}} \, \sin(\omega t)$
applied to the resistance $1/\alpha$. 

Equation (\ref{cinque}) and (\ref{sei}) where spatially discretized and
integrated in time with a fourth-order Runge-Kutta method.
The following numerical results were obtained for a junction with
$l=4$ and $\alpha=0.1$. This means that the junction is four times
longer than $\lambda_{\mathrm{J}}$, and that the current leakages are about
10\,\% with respect to the critical current, $I_{\mathrm{c}}$.

\begin{figure}[t]
\epsfig{file= 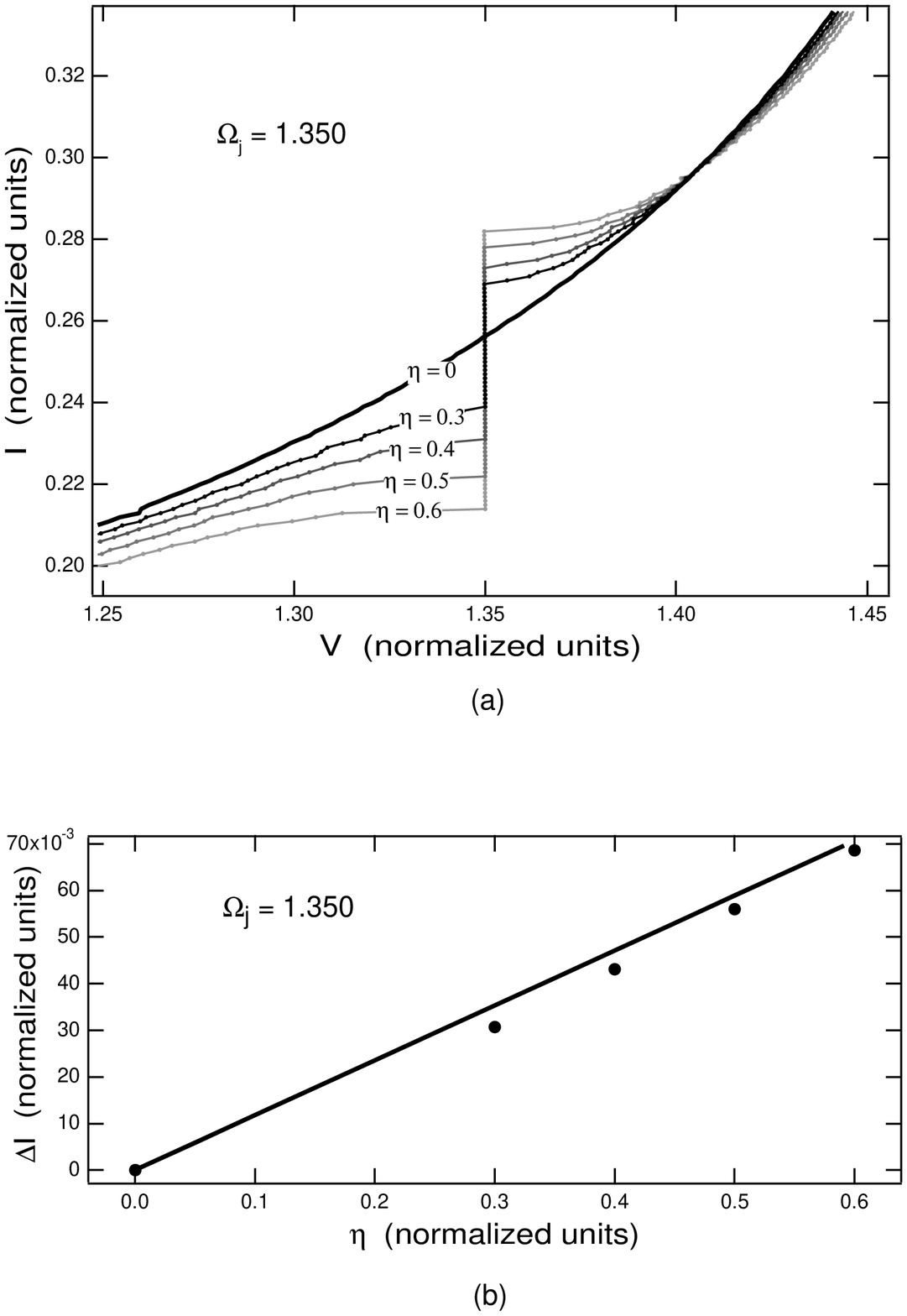,width=0.95\columnwidth}
\caption{
\small{(a) Calculated I-V curves for a junction with 
\protect{$\lambda_{\mathrm{J}}=4$},
\protect{$\Omega_{\mathrm{J}}=1.35$}, 
\protect{$\alpha=0.1$}, and for and different values of $\eta$.
(b) Dependence of the range of phase-locking interval upon
the rf current amplitude, $\eta$.
The locking takes place when the frequency of the external drive is
twice the fluxon oscillation frequency.
The straight line in (b) represents the equation 
(\protect{\ref{sette}}), and the dots are the heights of the
phase-locking steps evaluated from the I-V curve.
The agreement is very good.}
}
\label{LayoutOmega1350}
\end{figure}

The I-V curves in Figure \ref{LayoutOmega1350} were obtained
with a bias increment $\delta \gamma = 10^{-4}$.
For each increment, $\delta \gamma$, we let the routine run for
a transient of 1000 rf cycles to stabilise the fluxon
oscillations in the junction.
After this transient the average value
of the dc voltage, $\langle \phi_{t} \rangle$, was taken over 1024 
rf cycles.
The dc voltage across the junction is instantaneously given by
the second Josephson equation, that in our normalised units, is:
\begin{displaymath}
\phi_{t} = V.
\end{displaymath}
%

%-------------------------------------------------------------
%-------------------------------------------------------------

In Figure \ref{LayoutOmega1350}(a) the I-V curves obtained
from the system  (\ref{cinque})-(\ref{sei}) with the parameters 
given before, are shown.
We chose the bias point on the Zero Field Step corresponding
to the Josephson plasma frequency $\Omega_{\mathrm{J}}=1.35$. 
The phase-locking steps will then appear at
\begin{displaymath}
V = \langle \phi_{t} \rangle = \Omega_{J}.
\end{displaymath}

In Figure \ref{LayoutOmega1350}(b) we show a comparison between
the heights of the phase-locking steps as evaluated from the
Figure \ref{LayoutOmega1350}(a) (dots)
and the equation (\ref{sette}) (straight line).
The agreement is very good as previous works show
\cite{CirLloApr87, CirCocFeb94, CirRotJul95, GroSep92, SalSamMay89,
PedDavJan90}.

\section*{Conversion gain for a resonant Josephson mixer}
From the equations (\ref{due}) and 
(\ref{quattro}) we derive the following general expression
for the conversion efficiency in a Josephson effect mixer
as a function of the position $V$ on the I-V curve:
\begin{equation}
G(V)= \frac{R_{\mathrm{dyn}}(V, \eta)}{R(V)} 
\left[ \frac{\partial (\Delta I)}
{\partial I_{\mathrm{rf}}} \right]^{2},
\label{otto}
\end{equation}
where we assume the output coupling efficiency, $C_{\mathrm{if}}$,
to be unitary.
$R$ represents the shunt resistance to the junction,
and
\begin{equation}
\frac{1}{R_{\mathrm{dyn}}(V, \eta)} 
= \frac{\partial I(\gamma, \eta)}{\partial V}.
\label{rdyn}
\end{equation}

For a resonant Josephson effect mixer, from equations 
(\ref{quattro}) and (\ref{sette}),
we see:
\begin{equation}
\chi=
\frac{\partial (\Delta I)}{\partial I_{\mathrm{rf}}}=
\frac{\partial (\Delta I)}{\partial \eta}=
\frac{1}{\sqrt{2}} \, \alpha \, R_{\mathrm{p}},
\label{nove}
\end{equation}
i.e., that $\chi$ is a constant depending only on the bias point on
the Zero Field Step through the dynamical resistance $R_{\mathrm{p}}$
at the point where the phase-locking step will appear:
\begin{equation}
\frac{1}{R_{\mathrm{p}}} = { \left( \frac{\partial 
I(\gamma, \eta=0)}{\partial V}
\right) }_{V = \Omega_{\mathrm{J}}}.
\label{rp}
\end{equation}
This case is very interesting especially because for a non-resonant
Josephson mixer $\chi$ has a very complicate behaviour,
while the (\ref{nove}) is a simply linear relation.
This simplicity is a remarkable property of the resonant Josephson effect mixer
that we are studying in this paper.

In the equation (\ref{otto}) $R$ is the shunt resistance to the
junction which is the parallel resistance between the ohmic resistance, 
$1/\alpha$ and 
dynamic resistance of
the Zero Field Step\cite{CirRotJul95}, $R_{\mathrm{d}}(V)$, where
\begin{equation}
\frac{1}{R_{\mathrm{d}}(V)} = \frac{\partial I(\gamma, \eta=0)}{\partial V}.
\label{rd}
\end{equation}
The approximation made in Figure \ref{modCirillo}
is still valid because we look at a small part of the I-V curve around the
bias point.
We have then that:
\begin{equation}
\frac{1}{R} = \frac{1}{R_{\mathrm{d}}(V)} + \alpha \approx 
\frac{1}{R_{\mathrm{d}}(V)},  
\left( R_{\mathrm{d}}(V) \ll \frac{1}{\alpha} \right).
\label{dieci}
\end{equation}

The final expression for the conversion efficiency for a 
resonant Josephson effect mixer therefore is:
\begin{equation}
G=\frac{R_{\mathrm{dyn}}(V, \eta)}{R_{\mathrm{d}}(V)} \chi^{2}=
\frac{R_{\mathrm{dyn}}(V, \eta)}{R_{\mathrm{d}}(V)} \left[ \frac{1}{\sqrt{2}} 
\alpha R_{\mathrm{p}} \right]^{2}.
\label{undici}
\end{equation}

\section*{Results}
\begin{figure}[t]
\epsfig{file=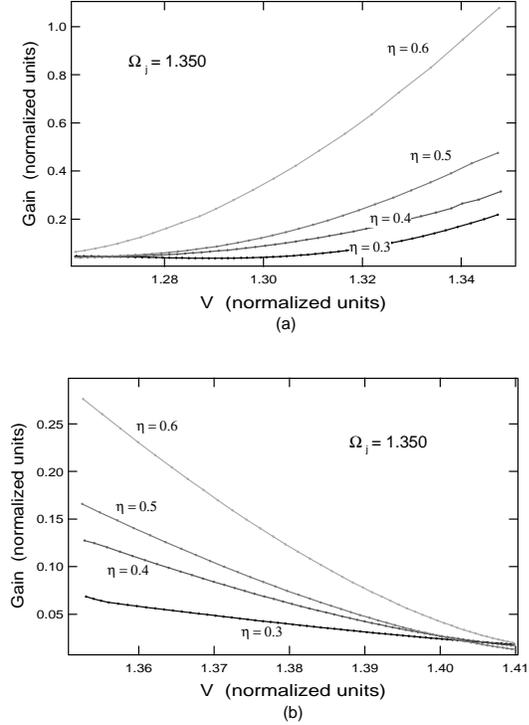,width=0.95\columnwidth}
\caption{
\small{Conversion efficiency for a resonant Josephson effect
mixer. The junction is driven by an external signal whose
current amplitude is given by $\eta$. In (a) we look at the
branch of I-V curve on the left with respect to the phase-locking
step. In (b)  we look at the branch of I-V curve
on the right  with respect to the phase-locking step.
We observe from (a) that it is possible to obtain a conversion
efficiency bigger than unitary for a bias point close to the
phase-locking step.}
}
\label{LayoutGain1350}
\end{figure}

In order to evaluate the conversion efficiency from equation (\ref{undici})
we have to calculate the dynamic resistances in that equation. 
Attention must be paid to this because an inaccurate estimate
of these parameters can give rise to big inaccuracy in the
conversion efficiency that we want to study in this paper.
We evaluated the dynamical resistances by differentiating the eighth-order
polynomial curve fitting of the I-V curve points spaced by a
$10^{-4}$ current step.
The results are shown in Figure \ref{LayoutGain1350}.
In Figure \ref{LayoutGain1350}(a) we look at the
branch of I-V curve on the left with respect to the phase-locking step.
In Figure \ref{LayoutGain1350}(b) we look at the branch of I-V curve
on the right  with respect to the phase-locking step.

%-----------------------------------------------------------
%----------------------------------------------------------

In Figure (\ref{LayoutGain1350}a)we find a conversion efficiency exceeding
units provided that the bias point is close enough to the phase-locking
step. Here we still assume, as mentioned before, an ideal rf-if coupling
($C_{\mathrm{if}}=1$), and an uniformly driven junction
(this can be achieved with an overlap-geometry
junction).

From Figure \ref{LayoutGain1350} we also observe that:

For $\Omega_{\mathrm{j}}$ and $\eta$ given, the conversion gain is larger
where the I-V curve is flatter. Indeed, there we
have a larger dynamic resistance, $R_{\mathrm{dyn}}(V, \eta)$, 
and consequently a larger gain.

For $\Omega_{\mathrm{j}}$ and $\eta$ given, the conversion gain is larger
for a bias on the branch of the I-V curve
to the left of the phase-locking step.
This is easily explained by observing that the I-V curve of the
junction is generally flatter (giving a bigger dynamical resistance,
$R_{\mathrm{dyn}}(V, \eta)$) on the left branch.

For $\Omega_{\mathrm{j}}$ given, 
the conversion gain is larger for larger values of
$\eta$. Again, a larger $\eta$ gives a flatter shape to the I-V curve.

\section*{Conclusions}
We have studied the conversion efficiency for a
resonant Josephson effect mixer
and we found a theoretical analytic expression for it. 
From the fundamental equations for a LJJ (long Josephson junction) 
we have obtained an estimate for the conversion efficiency:
it is possible to have a conversion efficiency exceeding units.

The necessary condition set by the sine-Gordon equation
is that the junction should be uniformly rf driven which can
be experimentally accomplished in the overlap geometry.
The calculations where also performed in the limit of an ideal
coupling between the junction and the if circuit ($C_{\mathrm{if}}=1$).

We conclude that a resonant Josephson effect mixer can have very good
performances.  
They can be explained as a consequence of the resonant nature
of the long Josephson junctions:
The linewidth of the signal is in fact very small
(1\,KHz at 10\,GHz, 1MHz at 100\,GHz \cite{CirRotJul95, ZhaWinApr94})
and this decreases the noise in the mixer.
The noise is in fact due mostly to down conversions of frequencies in the
linewidth of the mixer signal, and it decreases as the linewidth
shrinks:
Even very small signals can be detected with reasonably good conversion 
efficiencies.

\section*{Acknowledgments}
We would like to thank Matteo Cirillo for helpfull discussions 
about the fluxon dynamics and the phase-locking steps.

\end{document}